# SpliceCombo: A Hybrid Technique efficiently use for Principal Component Analysis of Splice Site Prediction

Srabanti Maji[1]

srabantiindia@gmail.com

Soumen Kanrar[2]

kanrars@acm.org

**Department of Computer Science and Engineering[1,2]**

**DIT University Mussorrie Diversion Road**

**Dehradun-248009, Uttarakhand, India**

## Abstract

The primary step in search of the gene prediction is an identification of the coding region from genomic DNA sequence. Gene structure in the case of a eukaryotic organism is composed of promoter, intron, start codon, exons, stop codon, etc. Splice site prediction, which separates the junction between exon and intron, though the sequence beside. The splice sites have huge preservation, however, the precision of the tool exhibits less than 90%. The main objective of this work to exhibits a hybrid technique that efficiently improves the existing gene recognition technique. Therefore to enhance the identification of splice sites, the respective algorithm needs to be improved. Over the last decade, the researcher paid more attention to improve the accuracy of a predicted model in this domain. Our proposed method, 'SpliceCombo' involves three stages. At initial stage, which considers the principal Component Analysis, based on the feature extracted. In the intermediate stage, i.e.,, the second stage Case- Based Reasoning is done, i.e., feature selection. The third stage uses support vector machine based along with polynomial kernel function for final classification. In comparison with other methods, the proposed SpliceCombo model outperforms other prediction models with respect to prediction accuracies. Particularly for donor splice site the methodology exhibits sensitivity is 97.25% accurate and specificity is 97.46% accurate. For acceptor Splice Site the sensitivity is 96.51% and Specificity is 94.48% correct.

# 1 Introduction

In the eukaryotes, it's very challenging to predict exon-intron structure in a sequence due to its complex structure and vast length. Research and analysis on the human genome show nearly 20,000–25,000 protein-coding gene exist [1]. There are nearly 100,000 genes in the human genome. Which indicates a huge number of genes are still unidentified [2], [3]. Most of the computational techniques achieve better performance, but inherent legacy with few drawbacks [4].

There are four different nucleotides in the DNA sequence i.e., A, C, G and T, in which group with three nucleotides symbolizes codon [5]. The eukaryotic genes coding sequence is separated by non-coding sequences, called as introns, which is absent in prokaryotes. An ORF is a DNA sequence commencing with a start codon (ATG) and terminating with the stop codon (TAA, TAG or TGA). The separating junction among the exon and intron is known as a donor splice site whereas, in case of an intron and an exon, it is called an acceptor splice site. The Figure 1 shows the central dogma of splice site [6].

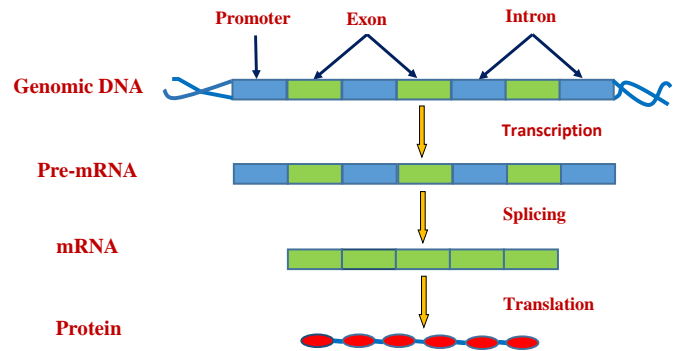

Figure 1: Central dogma of splice site

The splice site identification is performed by discriminating true from the rest of splice sites for both acceptor and donor sites. A wide range of tools has been designed based on the probabilistic approach [2]-[4], [7]-[11]. Those are support vector machine and neural network [8], [12]-[14], discriminant analysis based techniques [15], [16] and the information theory [17-18]. The non-linear transformation through neural networks (NN) and Support vector machines (SVM) describe the feature of adjoining di-nucleotides AG/GT.

Through the weight matrix method (WMM) many researchers have identified splice sites Arita et al., [17], adopted this technique in Net Plant Gene [19] and NNSplice [10]. A better accuracy was accomplished by first order Markov model (MM1, WAM) [20], [9]. Burge developed a tool Gen scan, it is maximal dependence decomposition (MDD) in the decision tree methodology [7]. It has been recommended that by combining signal/content methods, with other statistical tools, like MDD, WMM, MM1 etc., a noticeable improvement can be possible. GeneSplicer is combination of second order Markov models and MDD also a process under this category [15]. Rajapakse presented a hybrid tool that collaborates back propagation neural networks (BPNN) and MM2, its need large sequence windows [16]. A statistical technique like Principal component analysis (PCA) is most frequently used for making predictive models. Overton and Haas has described a case-based system, which uses grammars and features of genes for instance promoter regions and other signals [21]. Our proposed method consider PCA and CBR which combine with SVM to enhance the accuracy and efficiency of splice site prediction. It has been found that this proposed model i.e., SpliceCombo shows superior performance when analyzed with other present splice site prediction programs. The main objective of this work to exhibits a hybrid technique that efficiently improves the existing gene recognition technique.

## 2  Materials and methodology

### 2.1 Evaluation Datasets

To measure the performance of proposed algorithm, we have considered three datasets of the splice site. The detail description of the datasets is summarized as follows.

HS3D (Homo Sapiens Splice Sites data set), a dataset of intron, exons and splice sites, is our first considered dataset [20]. This Human genes dataset had been extracted from the GenBank and the length of each splice site sequence is 140bp. True donor and pseudo donor splice sites, which contained "GT" dinucleotides, are 2796 and 271937 in number respectively. Whereas true acceptor and pseudo acceptor sites, containing "AG" dinucleotides, are 2880 and 329374 number respectively. For acceptor splice site, AG is conserved at the index positions -69 and -70 of the sequences of the gene. Whereas for donor splice site, GT dinucleotide is conserved at positions -71 and -72 of the sequences. The ratio of the pseudo splice site with true splice site is brought as 10:1 this dataset is used to extract features for further modeling. The second dataset, DGSplicer [22], which is a true dataset, had been created by extracting 2381 member of true acceptor splice sites along with 2381 member of true donor splice sites from 462 distinguished human genes with multiple-exon [20]. After excluding two donors and one acceptor splice site from the set to create a group of 2380 true acceptor and 2379 true donor splice sites. A huge group of 283062 pseudo donor and 400314 pseudo acceptor sites are accumulated from 462 members of human genes, to form the imposter dataset. The donor splice site's window size is18 nucleotides in with the range {-9 to +9} among GT, at the locations +1 and +2. Whereas the same for the acceptor splice sites of 36 nucleotides are {-27 to +9} among AG at the locations -26 and -27. The third dataset is created with the *Drosophila* genomic sequence model. Which is trained in a dataset provided by the organizers of the GASP experiment [23] and comprised of DNA entries from to GenBank. The complete dataset consisted with of 275 members of multiple exons and member of 141 single exon *Drosophila* genes. Moreover for the purpose of training the codon usage Markov models, which is well-annotated gene structure data set and all available coding sequences from mRNA sequence. The entries in GenBank for *Drosophila melanogaster* are also used. The contiguous genomic *Adh* sequence is run against the non-redundant GenBank protein database. BDGP Web site provides a comprehensive analysis of this Adh region [24].

## 2.2 Overview of the Propose Model

The Process of splice site identification is segmented into two separate classification modules – donor site classification and acceptor site classification. Additionally, two modules consisting of three phases, those are assembled to recognize acceptor and donor splice sites. This model applies for various significant aspects i.e., features extraction, feature selection, and classification. The fundamental steps are as follows:

1. Feature extraction: Principal component analyses (PCA) are used to generate derived values or features projected to be informative and non-redundant, from the initial set of measured data.
2. Feature selection: Assessment of difference between every feature is carried out to select more enlightening features. This is performed by using case-based reasoning (CBR).
3. Classification: The Support Vector Machine Classifier with the polynomial kernel is trained on the probabilistic frameworks. Here Figure 2 depicts the proposed model's architecture.

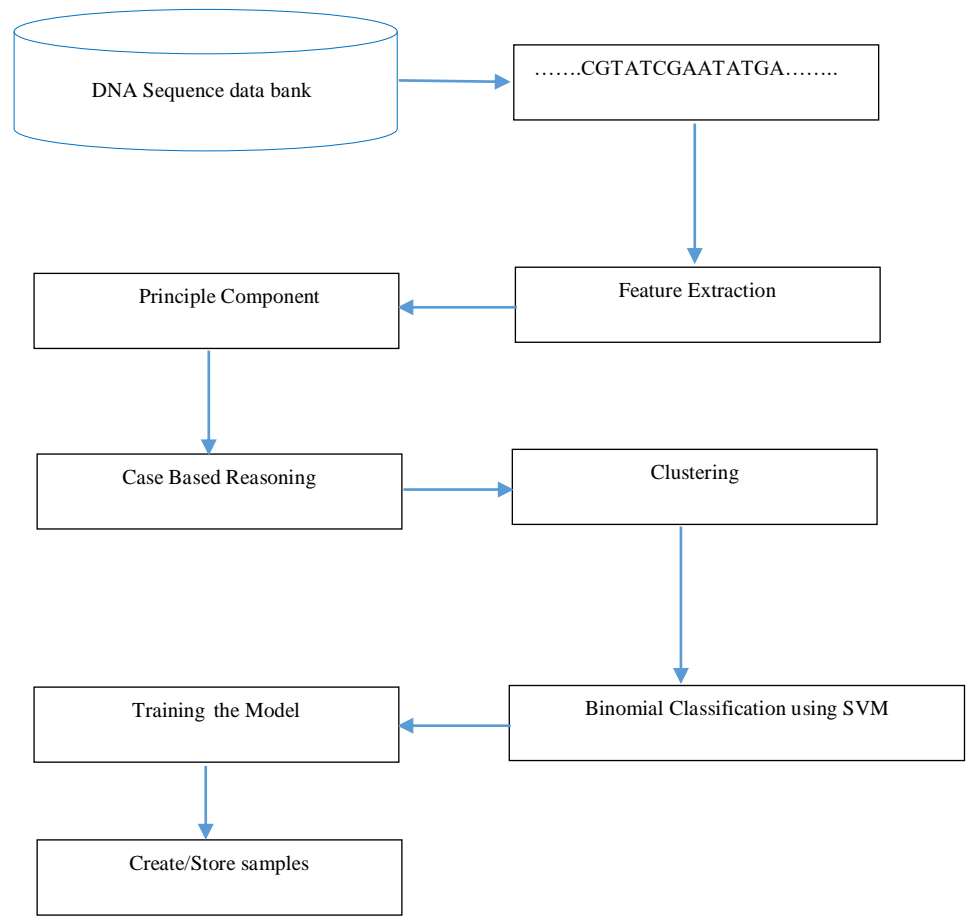

Figure 2: SpliceCombo Model

## 2.2.1 Feature Extraction

For the extraction of features from microarray data, principal component analysis (PCA) has been extensively used for the analysis of the image and speech data [25]. It is extensively used in various software packages [26], [27]. Figure 3 presents a two-variable data set with a solid ellipse, and mapped in the U-W coordinate. The U axis indicates the principal direction along the variation of data, while W axis is the other principal direction orthogonal to U. The data are transformed into (U, W) coordinate value for each (X, Y) coordinate data. By using principal directions of the variance (i.e., the U-W axis system according to Figure 3) the principal component analysis can find the axis system, for a given data set. The directions U and W are labeled as principal components.

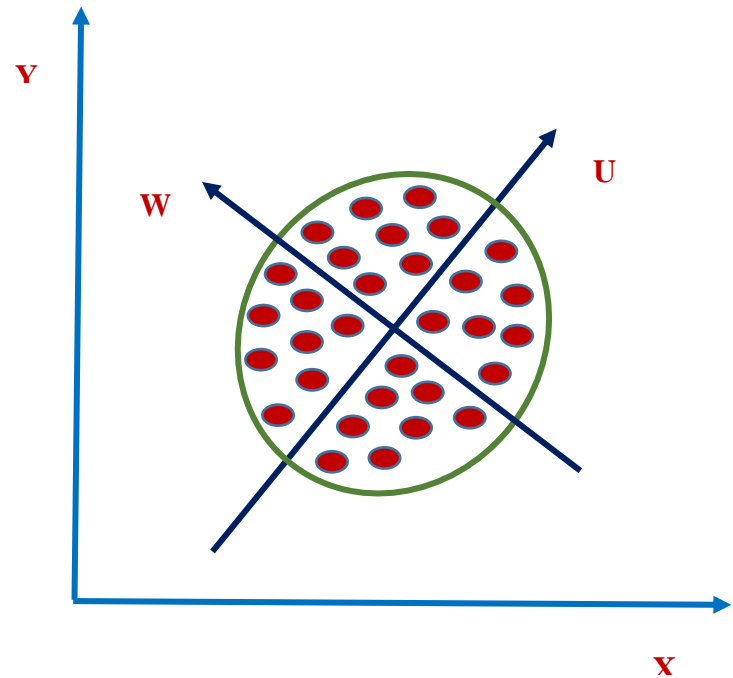

Figure 3: Representation of Principle Component Analysis for the given data.

It is possible to project from d dimensional gene expressions into a small number of Principal Components (i.e., PC). Let Z is the p index random vector such that Z= $[Z_1, Z_2, \ldots, Z_P]^T$. Here the matrix M represents the covariance of $Z_1, Z_2, \ldots, Z_p$.

$$M = \begin{bmatrix} M_{11} & M_{12} & \ldots & M_{1p} \\ M_{21} & M_{22} & \ldots & M_{2p} \\ . & . & . & . \\ . & . & . & . \\ M_{p1} & M_{p2} & M_{p3} & M_{pp} \end{bmatrix}$$

Now, the diagonal elements $M_{11}, M_{22}, \ldots, M_{pp}$ represent the variance for $Z_1, Z_2, \ldots, Z_p$. It is reflected on the *p* index variation degree. Therefore, the total variation degree of *p* index will be expressed as $M_{11} + M_{22} + \cdots + M_{pp}$. If we want to obtain a new index instead of the original *p* index, then new index will include the original information, and expressed according to

Y=AZ, here, $Y_i = a_{i1}Z_1 + a_{i2}Z_2 + \cdots + a_{ip}Z_p$

$$\begin{bmatrix} Y_1 \\ Y_2 \\ \vdots \\ Y_p \end{bmatrix} = \begin{bmatrix} a_{11} & .. & .. & a_{1p} \\ a_{21} & .. & .. & a_{2p} \\ \vdots & \vdots & \vdots & \vdots \\ a_{p1} & .. & .. & a_{pp} \end{bmatrix} \begin{bmatrix} Z_1 \\ Z_2 \\ \vdots \\ Z_p \end{bmatrix}$$

Assume that $\lambda_1 \geq \lambda_2 \geq \cdots \geq \lambda_\gamma$ ($\gamma \leq p$) are the non-zero value for the characteristic equation, $\det|A - \lambda I| = 0$. Then $M_{11} + M_{22} + \cdots + M_{pp} = \lambda_1 + \lambda_2 + \cdots + \lambda_\gamma$. So, we can retrieve the $\gamma$ on the whole index of $y_1, y_2, \ldots, y_\gamma$, whose variance is equal to the variance of *p* index array. The information transfers in the $\gamma$ index is equal to the original *p* index contains information. If *p is* much larger than γ, then (PCA) technique will reduce the index without affecting the result. Because of the overall index $Y_i = a_{i1}Z_1 + \cdots + a_{ip}Z_p$ is larger when the variance is $\lambda_i$, so the ability to synthesize the *p* index of $y_i$ is the strongest. The first, second… and the $\gamma^{th}$ principal components are defined by $y_1$, $y_2, \ldots, y_\gamma$, respectively. Now

$$\frac{\lambda_\gamma}{\lambda_1 + \lambda_2 + \ldots + \lambda_\gamma} = \frac{\lambda_\gamma}{M_{11} + M_{22} + \ldots + M_{pp}} \qquad (1)$$

In the above expression, the proportion of $y_\gamma$ variance in the total variance is the variance contribution rate along the $\gamma^{th}$ principal component [29].

### 2.2.2 Feature Selection

Feature selection has an important role in pattern classification for better interpretation of results. The Case-Based Reasoning (CBR) doesn't need an explicit domain model, it learns by procuring new knowledge as cases. So the maintenance of large volumes of information is easy [30].

A system of the CBR working cycle contains four REs (reference Figure 4).

I. RETRIEVE the most analogous cases.

II. REUSE the cases to resolve the current obstacle.

III. REVISE the projected solution if required.

IV. RETAIN the contemporary solution as a portion of that case.

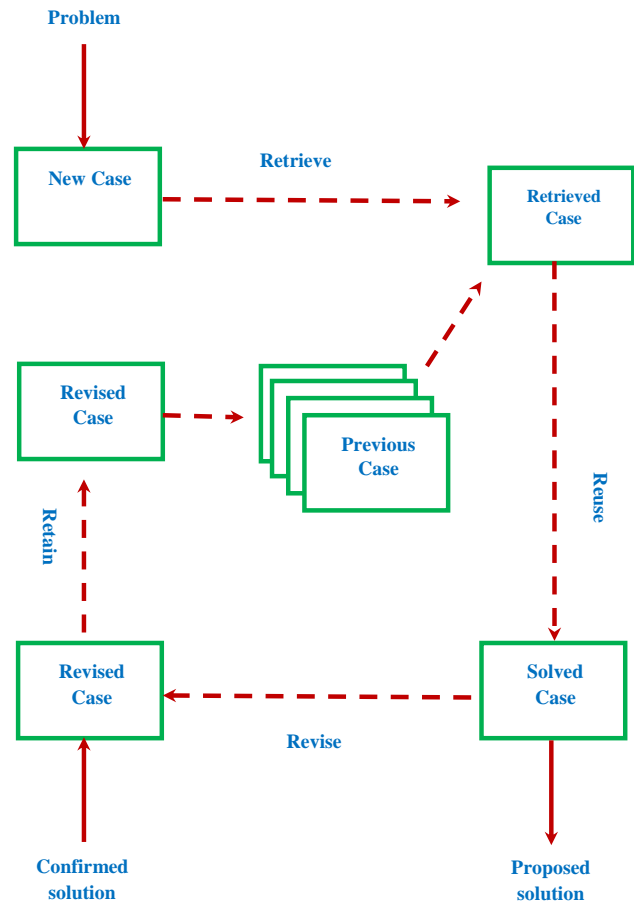

Figure 4: Case-Based Reasoning (CBR) cycle.

Costello and Wilson show the application of a CBR methodology for gene prediction, which performs the comparisons between DNA to isolate significant coding sections [32]. After selecting input variables through PCA, CBR is employed to compute the weightings for these variables. Then the cluster of these cases forming into groups. There are two phases in this reasoning:

**First Phase:**

Identify the minimum evaluation value E(v) by taking the following steps.

**Step 1:** Calculate Weight Distance Matrix. The initial value is produced randomly.

$$d_{pd}^{(v)} = d^{(v)}(e_p, e_q) = (\sum_{j=1}^{T} v_j^2 (x_{pj} - x_{qj})^2)^{\frac{1}{2}} = (\sum_{j=1}^{T} v_j^2 x_j^2)^{\frac{1}{2}} \quad (2)$$

$v_j$ : $weight$, $v_j \in [0,1]$, $j(1 \leq j \leq a)$
$T$: Total number of cases
$e_p$, $e_q$: case $p$; $q$;
$a$: All of the important factors

**Step 2:** After calculating $d_{pq}^{(v)}$ for each case first, then calculate the similar matrix $GN_{pq}^{(v)}$.

P =1,....,L,  $q \prec p$
$\alpha$ : User defined.

$$GN_{pq}^{(v)} = \frac{1}{1 + \alpha . d_{pq}^{(v)}} \quad (3)$$

**Step 3:** Calculate E(v)

$$E(v) = \frac{2.\left[\sum_{pq(q<p)} \sum \left(GN_{pq}^{(v)}\right)\left(1 - GN_{pq}^{(1)}\right) + GN_{pq}^{(1)}(1 - GN_{pq}^{(v)})\right]}{T.(T-1)} \quad (4)$$

$GN_{pq}^{(1)}$ : Weight equals 1 for each important variable.

**Step 4:** Use gradient descent technique to change the weighting of $\Delta v_j$ to minimize the value E(v)

$$\Delta v_j = -\eta \frac{\partial E}{\partial v_j} \tag{5}$$

$\eta$ : Learning stage

$$\frac{\partial E(v)}{\partial v_j} = \frac{2\left[\sum_{pq(q<p)}\sum \left(1 - 2 \times GN_{pq}^1\right) \times \frac{\partial GN_{pq}^{(v)}}{\partial d_{pq}^v} \times \frac{\partial d_{pq}^{(v)}}{\partial v_j}\right]}{T(T-1)} \tag{6}$$

$$\frac{\partial SM_{pq}^{(v)}}{\partial d_{pq}^v} = \frac{-\alpha}{\left(1 + \alpha \times d_{pq}^{(v)}\right)^2} \tag{7}$$

$$\frac{\partial d_{pq}^{(v)}}{\partial v_j} = \frac{v_j \left(x_{pj} - x_{qj}\right)^2}{\left(\sum_{j=1}^{a}\left(v_j^2 \left(x_{pj} - x_{qj}\right)^2\right)\right)^{\frac{1}{2}}} \tag{8}$$

**Second Phase:** Find the best clusters for the case
Step 1: Provide a threshold value β ϵ (0, 1).
Step 2: assume GN = $GN_{pq}^v$
Step 3: Initialize GN1=GN, GN = $s_{pq}$

$$s_{pq} = \max_k (\min(GN_{pk}^v, GN_{kq}^v)) \tag{9}$$

Step 4: When GN1 ⊂ GN then moves to step 5; if not, Assume GN = GN1, and then go back to the step 3
Step 5: Identify the clusters, If result of the Case q and Case p are equal while $S_{pq} \geq \beta$ then they will belong to the same group.

## 2.2.3 Classification

The data classification is the process of organizing data into multiple categories in more efficient way. Structural Risk Minimization (SRM) technique is an inductive principle used for the formulation of SVM [33], [34]. That is superior to the Empirical Risk Minimization (ERM) principle and used with neural networks. In case of SVM classifier, the generalization error gets reduced when the margin is high [35].Figure.5 shows the SVM with hyper-plane and margin.

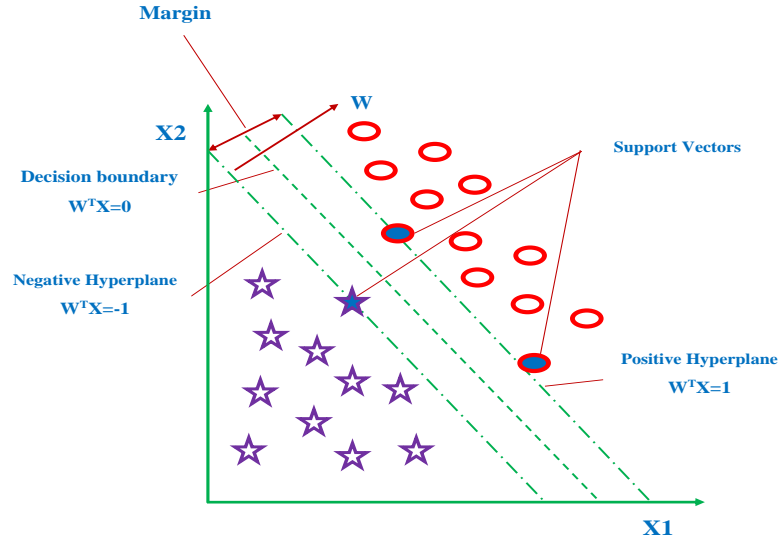

Figure 5: Support Vector Machine with hyper-plane and margin.

Dual formulation in SVM is obtained by using the Lagrange multipliers technique, expressed through variables $\alpha_i$. For solving the optimization problem, SVM classification is given by:

$$Maximize\ f(\alpha) = \sum_{I=1}^{C}\alpha_I - \frac{1}{2}\sum_{I=1}^{C}\sum_{J=1}^{C}\alpha_I\alpha_J B_I B_J K(A_I, A_J) \qquad (10)$$

$$\text{Subject to} \qquad \sum_{I=1}^{C} \alpha_I Y_J = 0, \ 0 \leq \alpha_I \leq H, I = 1,...C$$

In the above equation C, denotes the number of training data, $A$ is input vectors and $B$ defines the class value that can be either -1 or 1 and $H$ is trade-off parameter for performance generalization. Here f is the objective function, and $Y_i$ expression is achieved with the base 8 from above equation. The dual foundation stimulates the growth of the weight vector with reference to the input:

$$w = \sum_{I=1}^{C} B_I \alpha_I A_I \qquad (11)$$

In case of a soft-margin SVM distinct data points of $A_I$ in which $\alpha_I > 0$ are those positions that are within or on the margin, known as support vectors.

Assume that D is the given DNA sequence, the SVM classifies is trained on the decision function:

$$o(D) = sign\left[\sum_{I=T} \alpha_I y_I K(A_I, D)\right] \qquad (12)$$

Set of support vectors are represented by T.

For classification purpose, we have used Support vector Machine along with polynomial kernel function. According to the weighted sample on the non-parallel hyper-plane, the quadratic programming problems are given as follows:

$$K(A, D) = (<A \bullet D> + 1)^2 \qquad (13)$$

The Dot product is symbolized by $<\bullet>$.

Then equation (6) becomes

$$K(A, D) = \sum_{(I,J)=(1,1)}^{(m,m)} (A_I, A_J)(D_I, D_J) + \sum_{I=1}^{m} (\sqrt{2A_I})(\sqrt{2D_I}) + 1 \qquad (14)$$

In the above equation, the number of dimensions in vectors A and D are considered as m, $i^{th}$ element in vectors $A_I$ and $D_I$. After substituting equation (7) into (5), the output $O(D)$ becomes $2^{nd}$ order polynomial function over D. Therefore the sequence of length C, is a vector of conditional probabilities denoted by D:

$$D = \left[ P(Q_2|Q_1), P(Q_3|Q_2), ..., P(Q_C|Q_{C-1}) \right] \qquad (15)$$

## 2.3 Model Design

Splice site identification process is separated into two different modules i.e., acceptor site and donor site module. For each module different tools are produced, e.g., for HS3D donor data-set, one SpliceCombo tool is created and trained. The HS3D donor test dataset is utilized for assessing the classification performance of the model similarly. Another SpliceCombo tool is trained and tested with HS3D acceptor dataset. DGSplice and Drosophila, donor and acceptor datasets are used for training and testing purpose.

## 2.4 Model Learning

The proposed model's training is accomplished in three phases i.e., feature extraction using PCA, feature selection using CBR and the SVM with polynomial kernel training of degree 2. SpliceCombo model uses true and false splice site training data. The desired output for a result is fixed to +1 or -1 that depends on their class label. We have used MATLAB for implementing the support vector machine based classification [28], [36].

## 2.5 Model Comparison

For performance comparison among SpliceCombo and other models, we have selected closely related method. Support Vector Machine (SVM) with zero order Markov Model (MM0) was used for preprocessing performance comparison with our proposed models.

## 2.6 Performance Measures

This model's classification performance is assessed through the Receiver Operating Characteristic (ROC) curve that provides a degree of tradeoff between the true positive rate (TPR) and false positive rate (FPR). Percentage of accurate prediction of true splice sites is called sensitivity, $S_n$ while specificity, $S_p$ is referred to as the percentage of accurate prediction of pseudo splice sites. $S_N$ (TPR) and FPR, $S_P$ are shown through the below equations:

$$Sensitivity(S_N) = \frac{TP}{TP + FN} \qquad (16)$$

$$Specificity(S_P) = \frac{TN}{TN + FP} \qquad (17)$$

$$FPR = 1 - S_p = \frac{FP}{FP + TN} \qquad (18)$$

$$precision = \frac{TP}{TP + FP} \qquad (19)$$

FN->False Negative     TP-> True positive

TN->True negative      FP->False Positive

A true donor or true acceptor site is categorized as true donor or true acceptor site. Here TP it is known as a true positive. A false donor or false acceptor is wrongly anticipated as a true donor or true acceptor site, FP is known as a false positive. In similar a manner, the false donor or false acceptor site which is also categorized as a false donor or false acceptor site. As it is known as a true negative and if a true donor or true acceptor site is wrongly categorized as a false donor or false acceptor. It is recognized as a false negative as specified in Table 1.

**Table 1:** Interpretations of TP, TN, FP and FN.

|  | Predicted positive | Predicted negative |
|---|---|---|
| Actual positive | True positives, TP | False negatives, FN |
| Actual negative | False positives, FP | True negatives, TN |

Accuracy ($ACC$) is the amount of the candidate sites which are correctly classified in the test data set by using the proposed SpliceCombo tool; it was evaluated through the following formula:

$$ACC = \frac{TN + TP}{TN + TP + FN + FP} \qquad (20)$$

Matthews's correlation coefficient (MCC) is the correlation coefficient within the observed and predicted binary classifications. It is outlined by the formula.

$$MCC = \frac{TP \times TN - FP \times FN}{\sqrt{(TP + FP)(TP + FN)(TN + FP)(TN + FN)}}$$

Where MCC returns a value between -1 and 1. Here completely well-trained classifiers have the value 1.

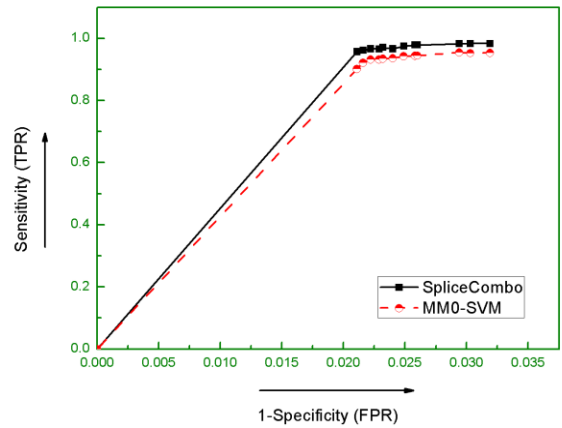

Figure 6: Performance comparison among MM0-SVM and SpliceCombo using HS3D donor dataset.

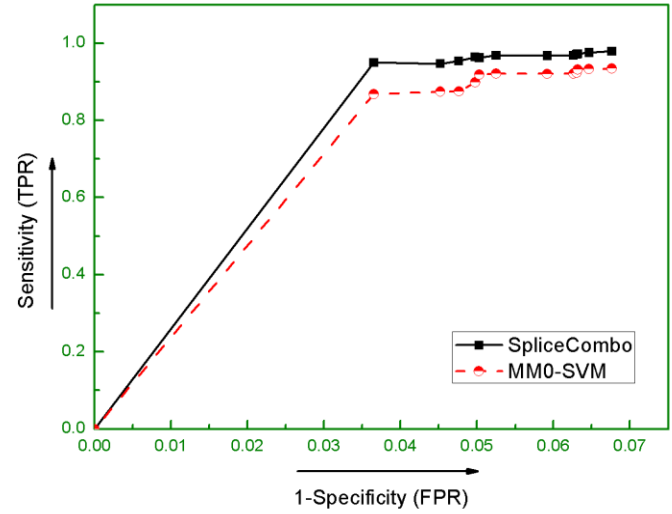

Figure 7: Performance comparison among MM0-SVM and SpliceCombo using HS3D acceptor dataset.

## 2.7 ROC Analysis

The receiver operating characteristic (ROC) curve is most popularly used to exhibits the performance of a binary classification [37]. The graph is plotted between 1-specificity (x-axis) vs. sensitivity (y-axis). The Euclidean matrix is used for approximating TPR as well as FPR. The more specific model represents the curve that approximates to (0, 0) point; (refer to Figure 6 to Figure13).

## 2.8 Cross Validation

Cross Validation technique is applied to enhance the performance of the predictive model. In the case of $x$-fold cross-validation, the authentic sample is partitioned into equal size subsamples along $x$. For testing the model, on each occasion one of the $x$ subsamples is used for test data set. While the other $x – 1$ subsamples are combined to form training data. To identify the prediction accuracy and compare the performance of SpliceCombo system with the other available methods, we considered twelvefold cross-validation (CV) technique [38], [39].

# 3 Results and Discussion

## 3.1 Selection of the Best Preprocessing Method

Methods like MM0 with SVM classifier have been used for preprocessing method selection. SpliceCombo tool has been used for splice site prediction. For accuracy comparison of MM0-SVM and SpliceCombo methods, HS3D donor and acceptor dataset have been used. Figure **6** and **7** present the ROC analysis for the models MM0-SVM and SpliceCombo respectively. This model is mainly used for splice site identification.

## 3.2 Comparison in the Predictive Performance

The proposed model yields 12-fold cross validation results based on, sensitivity ($S_n$), specificity ($S_p$), FPR and MCC for donor SpliceCombo. The acceptor SpliceCombo using HS3D are reported in Tables 2 and Table 3 respectively.

Table 2: Performance of Donor SpliceCombo with Polynomial Kernel Degree 2 for Identifying Donor (5' Splice) Sites

| S. No. | No of True Donor | No of Pseudo Donor | TP | FP | TN | FN | Sensitivity (Sn) | Specificity (Sp) | FPR | MCC |
|---|---|---|---|---|---|---|---|---|---|---|
| 1 | 233 | 22670 | 223 | 479 | 22191 | 10 | 0.95708 | 0.97887 | 0.0211 | 0.5448 |
| 2 | 235 | 22700 | 226 | 491 | 22209 | 9 | 0.96170 | 0.97837 | 0.0216 | 0.5439 |
| 3 | 240 | 23001 | 232 | 511 | 22490 | 8 | 0.96666 | 0.97777 | 0.0222 | 0.5427 |
| 4 | 235 | 22810 | 227 | 523 | 22287 | 8 | 0.96595 | 0.97707 | 0.0229 | 0.5339 |
| 5 | 241 | 22849 | 234 | 532 | 22317 | 7 | 0.97095 | 0.97671 | 0.0232 | 0.5377 |
| 6 | 238 | 23004 | 230 | 554 | 22450 | 8 | 0.96638 | 0.97591 | 0.0240 | 0.5254 |
| 7 | 237 | 22760 | 231 | 568 | 22192 | 6 | 0.97468 | 0.97504 | 0.0249 | 0.5237 |
| 8 | 236 | 22850 | 231 | 590 | 22260 | 5 | 0.97881 | 0.97417 | 0.0258 | 0.5176 |
| 9 | 238 | 23576 | 233 | 613 | 22963 | 5 | 0.97899 | 0.97399 | 0.0260 | 0.5121 |
| 10 | 235 | 22891 | 231 | 674 | 22217 | 4 | 0.98297 | 0.97055 | 0.0294 | 0.4931 |
| 11 | 238 | 22779 | 234 | 690 | 22089 | 4 | 0.98319 | 0.9697 | 0.0303 | 0.4910 |
| 12 | 237 | 22548 | 233 | 721 | 21827 | 4 | 0.98312 | 0.96802 | 0.0319 | 0.4817 |
| **Average Value** | | | | | | | **0.9725** | **0.9746** | **0.0253** | **0.5206** |

Table 3

Performance of Acceptor SpliceCombo with Polynomial Kernel Degree 2 for Identifying Acceptor (3' Splice) Sites

| S. No. | No of True Acceptor | No of Pseudo Acceptor | TP | FP | TN | FN | Sensitivity (Sn) | Specificity (Sp) | FPR | MCC |
|---|---|---|---|---|---|---|---|---|---|---|
| 1 | 240 | 27500 | 228 | 1006 | 26496 | 12 | 0.95 | 0.96349 | 0.03650 | 0.4103 |
| 2 | 246 | 27820 | 233 | 1258 | 26562 | 11 | 0.94715 | 0.95478 | 0.04521 | 0.3765 |
| 3 | 240 | 27450 | 229 | 1307 | 26143 | 11 | 0.95416 | 0.95238 | 0.04761 | 0.3671 |
| 4 | 250 | 28000 | 241 | 1393 | 26607 | 9 | 0.964 | 0.95025 | 0.04975 | 0.3667 |
| 5 | 239 | 27780 | 230 | 1398 | 26382 | 9 | 0.96234 | 0.94967 | 0.05032 | 0.3585 |
| 6 | 256 | 28300 | 248 | 1486 | 26814 | 8 | 0.96875 | 0.94749 | 0.05250 | 0.3616 |
| 7 | 248 | 28670 | 240 | 1697 | 26973 | 8 | 0.96774 | 0.94080 | 0.05919 | 0.3351 |
| 8 | 253 | 27600 | 245 | 1728 | 25872 | 8 | 0.96837 | 0.93739 | 0.06260 | 0.3349 |
| 9 | 244 | 27676 | 237 | 1745 | 25931 | 7 | 0.97131 | 0.93694 | 0.06305 | 0.3291 |
| 10 | 254 | 27855 | 247 | 1760 | 26095 | 7 | 0.97244 | 0.93681 | 0.06318 | 0.3341 |
| 11 | 249 | 27650 | 243 | 1788 | 25862 | 6 | 0.97590 | 0.93533 | 0.06466 | 0.3298 |
| 12 | 241 | 27554 | 236 | 1864 | 25690 | 5 | 0.97925 | 0.93235 | 0.06764 | 0.3197 |
| **Average Value** | | | | | | | **0.96512** | **0.94481** | **0.05519** | **0.3519** |

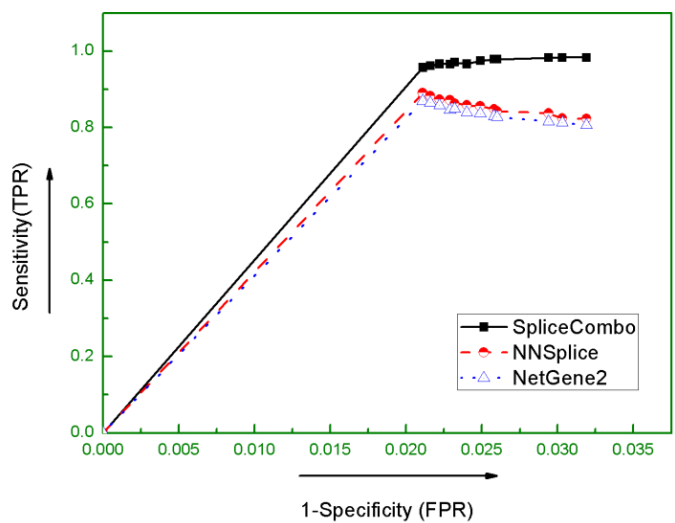

**Figure 8** Performance comparison of methods for table2.

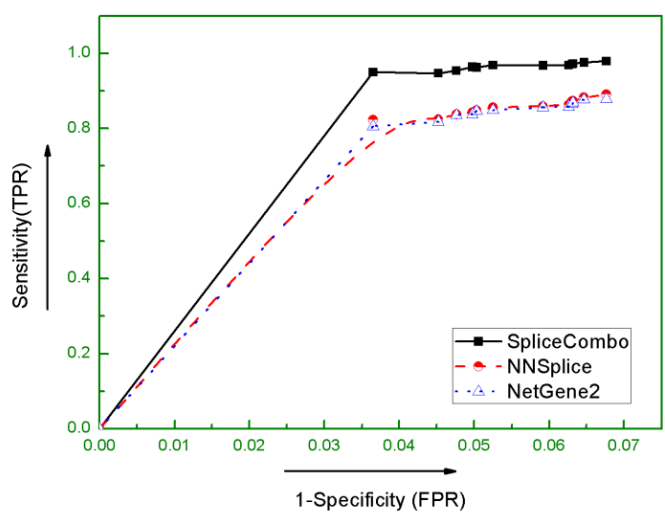

**Figure 9**     Performance comparison of methods for table3.

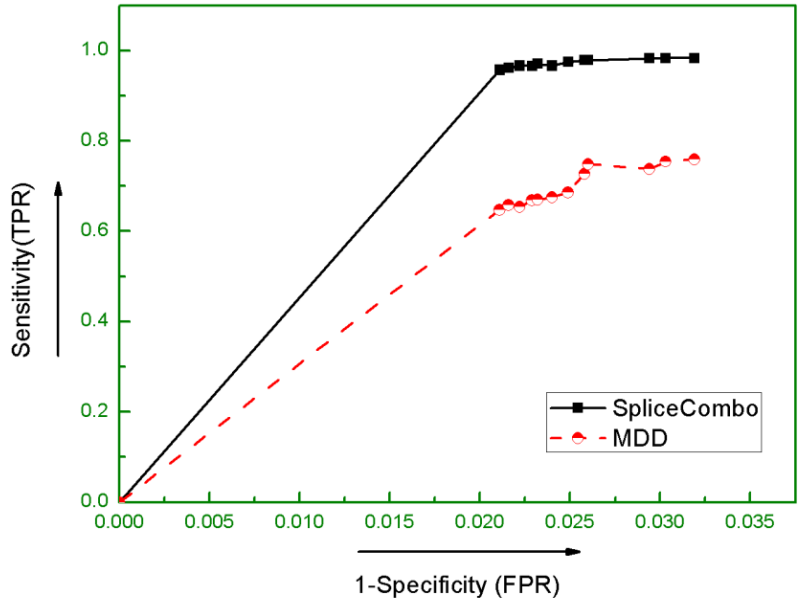

**Figure10:** Predictive performance comparison among the MDD and SpliceCombo are shown using DGSplicer donor dataset.

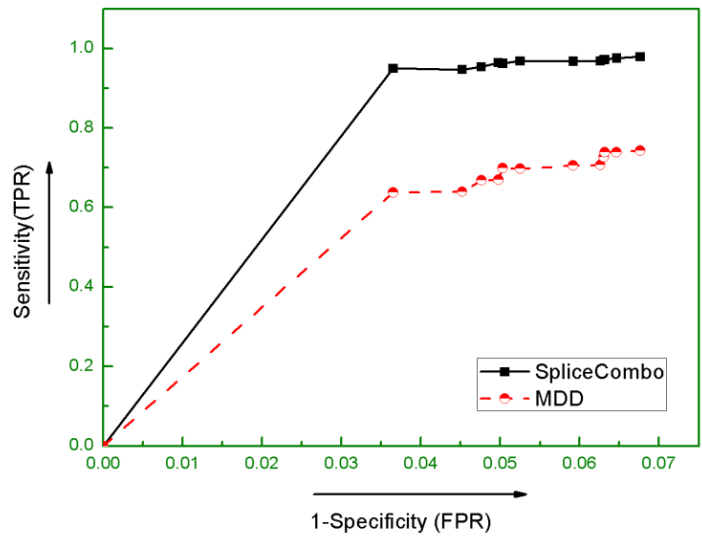

**Figure11:** Predictive performance comparison among the MDD and SpliceCombo are shown using DGSplicer acceptor dataset.

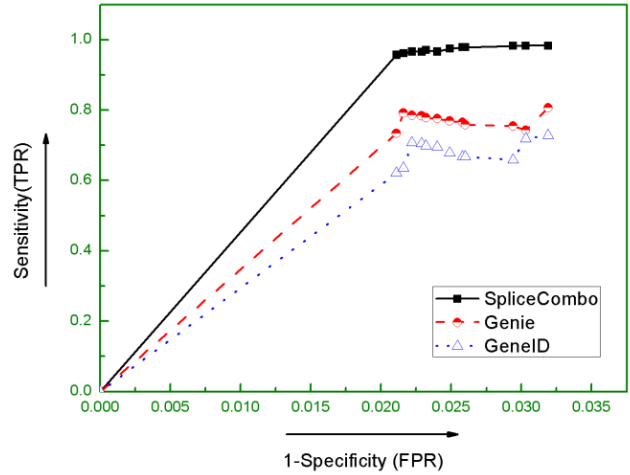

**Figure 12:** Predictive performance comparison among Genie, Geneid and SpliceCombo are shown using Drosophila donor dataset.

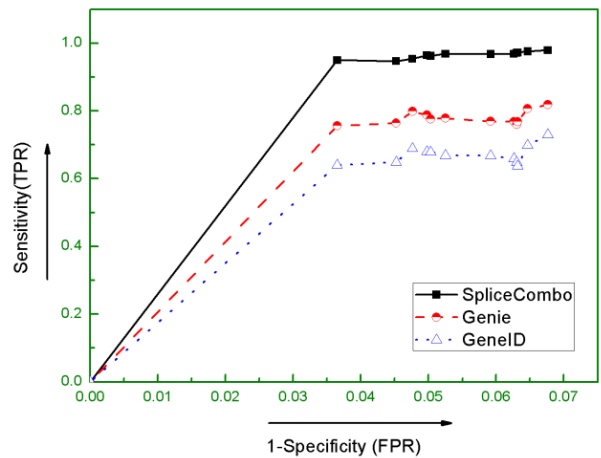

**Figure 13:** Predictive performance comparison among Genie, Geneid and SpliceCombo are shown using Drosophila acceptor dataset.

The comparison of performance between the SpliceCombo, NNSplice is done from the available dataset (http://www.frutfly.org/seq_tools/splice.html) and NetGene2 is trained through the available dataset (http://genome.cbs.dtu.dk/services/NetGene2/) and HS3D. For comparison, the standard TPR ($S_n$) and FPR(1- $S_p$) are used and monitored. The SpliceCombo exhibits superiority for acceptor and donor splice site identification.

Whereas second best performance been observed through NetGene2, as presented in Figure **8**, and Figure **9**. The maximum $S_n$ and $S_p$ values for SpliceCombo are 97.25% and 97.46% respectively for the donor splice site and 96.51% and 94.48% for acceptor splice site prediction. MDD method and DG Splicer dataset is used for comparison and validation for the performance of the proposed method. Here, SpliceCombo showed superior performance as exhibited in Figure10, Figure 11.

Further, we have used Drosophilla dataset for splice site detection accuracy comparison for the SpliceCombo method. Their performance measured with geneid and genie tool, are observed that SpliceCombo gives superior performance, as shown in Figure.12 and Figure 13.

# 4 Conclusion

In this work, we have analyzed a hybrid SpliceCombo system which identifies the features to predict a splice site junction. We have 12-fold cross-validation experiment for the verification of the results. This model correctly identifies a maximum of up to 97.25% of the true donor, 97.46% regarding the false donor and 96.51% of the true acceptor, 94.48% of the false acceptor splice sites. Furthermore, this method is simpler, more effective and usable to predict splice site junction on the enormous scale.